# Unravelling the phonon scattering mechanism in Half-Heusler alloys ZrCo$_{1-x}$Ir$_x$Sb (x = 0, 0.1, and 0.25)

Kavita Yadav[1], Saurabh Singh[2], Omprakash Muthuswamy[2], Tsunehiro Takeuchi[2], and K. Mukherjee[1]

[1]School of Basic Sciences, Indian Institute of Technology, Mandi, Himachal Pradesh-175005, India

[2]Research Centre for Smart Energy Technology, Toyota Technological Institute, Nagoya, 468-8511, Japan

Insight about the scattering mechanisms responsible for reduction in the lattice thermal conductivity ($\kappa_L$) in Half-Heusler alloys (HHA) is imperative. In this context, we have thoroughly investigated the temperature response of thermal conductivity of ZrCo$_{1-x}$Ir$_x$Sb (x = 0, 0.1 and 0.25). For ZrCoSb, $\kappa_L$ is found to be ~ 15.13 W/m-K at 300 K, which is drastically reduced to ~ 4.37 W/m-K in ZrCo$_{0.9}$Ir$_{0.1}$Sb. This observed reduction is ascribed to softening of acoustic phonon modes and point defect scattering, on substitution of heavier mass. However, no further reduction in $\kappa_L$ is observed in ZrCo$_{0.75}$Ir$_{0.25}$Sb, because of identical scattering parameter. This has been elucidated based on the Klemen's Callaway model. Also, in the parent alloy, phonon-phonon scattering mechanism plays a significant role in heat conduction process, whereas in Ir substituted alloys, point defect scattering (below 500 K) and phonon-phonon scattering (above 750 K) are the dominant scattering mechanisms. The minimum $\kappa_L$ is found to be ~ 1.73 W/m-K (at 950 K) in ZrCo$_{0.9}$Ir$_{0.1}$Sb, which is the lowest reported value till now, for n-type Zr based HHA. Our studies indicate that partial substitution of heavier mass element Ir at Co-site effectively reduces the $\kappa_L$ of n-type ZrCoSb, without modifying the nature of charge carriers.

Half-Heusler alloys (HHA) have been extensively investigated in the past few decades due to their exceptional properties such as large power factor, good mechanical and thermal reliability, and non-toxicity [1-5]. The devices fabricated out of these materials are useful due to their applicability in automobile industries, home heating appliances, body heat driven wrist watches, radio-isotope thermoelectric generators [6-7]. However, a major drawback in HHA is their high thermal conductivity ($\kappa$). To reduce $\kappa$, different effective mechanisms have been reported to increase the phonon scattering events: such as grain boundary scattering, phase separation, isoelectronic substitution, soft phonon modes, and reduction in grain size or introducing the nanostructures within crystal lattice [8-10]. For any system, $\kappa$ is the sum of thermal conductivities arising due to electronic ($\kappa_e$) and lattice ($\kappa_L$) degrees of freedom. In HHA, $\kappa_e$ is related to carrier concentration (n) and $\kappa_L$ is the independent parameter, which can be tuned by above mentioned strategies. Among these, the iso-electronic heavier element alloying in HHA can effectively decrease the $\kappa_L$ without affecting the charge disorder in the lattice. This decrement is observed due to point defect impurity scattering and modulation of phonon dispersion relation, like, reduction in group velocity of phonons and increment in Umklapp processes [11]. In an ideal crystal, an infinite $\kappa_L$ is observed at all temperatures due to absence of imperfections which inhibit the phonon-phonon interactions. Whereas, due to presence of crystal defects and lattice distortion a finite $\kappa_L$ is observed in real crystalline solid. Investigations of different phonon scattering mechanisms that are responsible for the reduction of $\kappa_L$ and transport of heat within these solids, remains an active area of research. In this context, XCoSb (X = Ti, Zr, Nb and Hf) HHA with VEC ~ 18 have been extensively explored with n and p type substitution at X, Co and Sb-site [12-14]. In ZrCoSb$_{1-x}$Sn$_x$ alloys, Yuan *et al.* has reported a drastic decrement in $\kappa_L$ due to strong phonon phonon scattering on incorporation of Sn at Sb-site [15]. The lowest value of $\kappa_L$ ~ 3.47 W/m-K at 973 K was obtained for ZrCoSb$_{0.7}$Sn$_{0.3}$. Similarly, Hsu *et al.* has demonstrated that substitution of p-type Fe at Co-site suppresses the $\kappa_L$ of ZrCo$_{1-x}$Fe$_x$Sb series [16]. Recently, D. Zhao *et al.* investigated Ni doped ZrCoSb and found that acoustic phonon and point defect scattering leads to significant decrement in $\kappa_L$, reaching ~ 5.52 W/m-K at 850 K in ZrCo$_{0.88}$Ni$_{0.12}$Sb [17]. He *et al.* has also studied the thermoelectric behaviour of Hf$_x$(ZrTi)$_{1-x}$CoSb$_{0.8}$Sn$_{0.2}$ and have attained the lowest $\kappa_L$ ~ 3.25 W/m-K in Hf$_{0.19}$Zr$_{0.76}$Ti$_{0.05}$CoSb$_{0.8}$Sn$_{0.2}$ at 973 K [18]. However, there are few studies on reduction of $\kappa_L$ in n-type XCoSb HHA through iso-electronic alloying, without modulating the nature of charge carriers.

Hence, in the present work, we have investigated the effect of iso-electronic Ir substitution at Co-site on the thermal transport behaviour of polycrystalline ZrCo$_{1-x}$Ir$_x$Sb (x = 0, 0.1 and 0.25). A detailed analysis has been carried out to understand the phonon scattering mechanisms in each alloy through both experimental and theoretical tools. It is observed that substitution of 10% Ir, suppresses the $\kappa_L$ significantly. However, due to similar effect of point defect on $\kappa_L$, further reduction in the magnitude of $\kappa_L$ is not observed in ZrCo$_{0.75}$Ir$_{0.25}$Sb. Interestingly, the lowest $\kappa_L$ ~ 1.73 W/m-K is attained in ZrCo$_{0.9}$Ir$_{0.1}$Sb near 950 K, which is so far lowest reported value in n-type Zr based HHA.

Polycrystalline alloys of ZrCo$_{1-x}$Ir$_x$Sb (x = 0, 0.1, and 0.25) are prepared using arc melting technique. The obtained ingots are crushed into powders using mortar and pestle.

Eventually, the powders are pressed into pellets and sintered using spark plasma sintering (SPS) technique at 1273-1473 K for 5 min under 50 MPa in an Ar flow atmosphere. To determine crystal structure and phase purity, room temperature (RT) X-ray diffraction (XRD) measurements are performed using rotating anode Bruker D8 Advance in Bragg-Brentano geometry (Cu-Kα; λ=1.5406 Å). Electron probe micro analysis (EPMA) (Make: JEOL JXL-8230) is used to confirm the composition of each alloy. Thermal conductivity (κ) measurements are done by using laser flash method (NETZSCH LFA 457). Seebeck coefficient (*S*) is measured by the steady state method [19]. Density of respective alloys is determined by using the Archimedes principle and are found comparable with the theoretical density. The phonon calculations are carried out using the constant-force method. A conventional unit cell and 2 X 2 X 2 superlattice are considered for phonon dispersion and density of states (DOS) calculations in the parent alloy as well as in the doped systems. The real space force constants are calculated using density functional perturbation theory (DFPT) as implemented in VASP. To perform these calculations, generalized gradient approximation in the form of Perdue–Burke–Ernzerhof is used. The phonon frequencies are determined using PHONOPY code [20] and a 16 X 16 X 16 mesh is considered for phonon DOS.

Fig. 1 represent the RT XRD patterns of $ZrCo_{1-x}Ir_xSb$ (x = 0, 0.1 and 0.25). The obtained patterns are analysed using Rietveld refinement method using FULL PROF software. The estimated lattice parameters are listed in table 1 and the remaining parameters are given in table S1 of the supplementary information. From the XRD patterns, it can be concluded that all alloys are formed in single-phase and crystallize in cubic structure (space group: *F-43m*). Here, we would like to mention that the compositions beyond x = 0.25 did not yield a crystallographically single phase alloy. From the table 1, it can be inferred that the lattice parameter (*a*) increases gradually with increment in Ir content at Co-site. It is also reflected in the shifting of peak towards lower angle side with increment in Ir concentration (as shown in inset of Fig. 1(b)). The observed trend is in accordance with the Vegard's law. This trend is observed due to the larger atomic radius of Ir in comparison with Co. The obtained lattice parameters agree well with the theoretical lattice parameters used for calculations. In HHA, one can determine the degree of disorder from intensity of (111), (200) and (220) diffraction lines [21]. The theoretical and experimental intensity ratio of $I_{111}/I_{220}$ and $I_{200}/I_{220}$ corresponding to each alloy is given in table S2 of the supplementary information. In ZrCoSb and $ZrCo_{0.9}Ir_{0.1}Sb$, the obtained intensity ratio is comparable with the theoretical calculated intensity ratio. However, in $ZrCo_{0.75}Ir_{0.25}Sb$, the experimental $I_{111}/I_{220}$ is found to be higher than theoretical intensity ratio, implying the presence of anti-site disorder, in this alloy. For determination of the nature of this disorder, further measurements like neutron scattering or high-resolution synchrotron measurements are required. From EPMA analysis, the average chemical composition corresponding to ZrCoSb, ZrCo$_{0.9}$Ir$_{0.1}$Sb, and ZrCo$_{0.75}$Ir$_{0.25}$Sb are determined as Zr$_{0.95}$Co$_{1.02}$Sb$_{1.03}$, Zr$_{0.98}$Co$_{0.90}$Ir$_{0.1}$Sb$_{1.02}$ and Zr$_{0.99}$Co$_{0.77}$Ir$_{0.24}$Sb$_{1.00}$, respectively.

Fig. 2 (a) shows the temperature response of κ of this series. For the parent alloy ZrCoSb, the RT κ is found to be ~ 15.13 W/m-K, which is less than the previously reported values [15-17]. In addition to this, a decreasing trend in the magnitude of κ with increment in temperature is observed. Interestingly, with partial substitution of heavier element Ir at Co-site i.e., in ZrCo$_{0.9}$Ir$_{0.1}$Sb, κ at RT decreases by 73%. A minimum value of ~ 1.73 W/m-K is attained at 950 K, which is among the lowest value reported till now for n-type ZrCoSb. With further increment in Ir concentration, i.e., in ZrCo$_{0.75}$Ir$_{0.25}$Sb, no significant change in RT κ is noted. This indicates that different scattering mechanisms are responsible for the reduced magnitude of κ in ZrCo$_{0.9}$Ir$_{0.1}$Sb and ZrCo$_{0.75}$Ir$_{0.25}$Sb. The broad hump noted in Ir substituted alloys in the 500-750 K temperature regimes is ascribed to the presence of bipolar effect. Similar feature is also noted in YNiBi [22], TiFe$_{0.5}$Ni$_{0.5}$Sb [23], Hf$_x$Zr$_{1-x}$NiSn$_{0.99}$Sb$_{0.01}$[24]. Here, we would also like to mention that substitution of Ir at Co-site does not alter the nature of charge carrier. This statement is substantiated from the observation of the negative value Seebeck co-efficient (*S*) across the series (Fig. S1 of supplementary information).

In solids, the heat transport mechanism is governed by the movement of charge carriers (κ$_e$) and lattice vibrations (κ$_L$). Here, κ = κ$_e$ + κ$_L$. Contributions of each component is extracted from κ by subtracting the individual component from it. The electronic contribution from κ is extracted using $\kappa_e = \frac{LT}{\rho}$, where L is the Lorentz number and *ρ* is the electrical resistivity of the alloy measured using the four-probe method [19]. As the members of this series are non-degenerate semiconductors, hence, variable values of L are used, and are defined as [25] $L = \left[1.5 + exp^{-\left[\frac{|S|}{116}\right]}\right]$. The temperature response of κ$_e$ for respective alloys is shown in the Fig. 2 (b). In ZrCoSb, the value of κ$_e$ at 300 K is estimated to be around 0.008 W/m-K, which also increases with temperature, reaching a maximum of ~ 0.04 W/m-K near 950 K. Similar trend in κ$_e$ is noted for Ir substituted alloys. In order to obtain κ$_L$ for respective alloys, the κ$_e$ is subtracted from κ. It can be observed from the Fig. 2 (c) that in all alloys there is significant contribution from the lattice vibrations in comparison to the motion of charge carriers. Also, in spite of the large mass difference between Ir and Co, negligible decrement in $\kappa_L$ of ZrCo$_{0.75}$Ir$_{0.25}$Sb is noted. This contradicts the fact that significant difference of atomic mass and radius between doped and host atom yield stronger phonon scattering. Similar observations are also noted in other HHA like, n-type NbCo (Sn,Sb), p-type (Nb,Zr)FeSb, and p-type Ti(Co,Fe)Sb [26-28].

In the subsequent paragraphs the role of different phonon mechanisms responsible for the observed magnitude and temperature dependent behaviour of $\kappa_L$ of this series of alloys has been discussed. Generally, the intrinsic phonon-phonon interaction is the dominant phonon scattering process at high temperatures, and in this case the $\kappa_L$ is given by [29]

$$\kappa_L = \frac{(6\pi^2)^{\frac{2}{3}} M}{V^{\frac{2}{3}} 4\pi^2 \gamma^2 T} v_g^3 \ldots\ldots (1)$$

where M is the average atomic mass of the alloy, V is the atomic volume, $\gamma$ is the Gruneisen parameter, $v_g$ is the average group velocity. The substitution of Ir at the Co-site alters the chemical composition and bonding between various atoms in the unit cell. This can lead to modifications in the $v_g$, which in turn can modify the strength of three phonon process. According to equation (1), $\kappa_L$ varies as $v_g^3$ which implies that the reduction of $v_g$ can suppress the $\kappa_L$. Similar mechanism was reported for the significant reduction of $\kappa_L$ in ZrCoBi based HHAs [30].

In order to investigate the applicability of this mechanism for the observed reduction in $\kappa_L$ in this series, the phonon dispersion and vibrational density of states are calculated. The calculations are performed on the conventional unit cell. All the phonon frequencies are found to be positive, which suggests that the structures models of Zr$_4$Co$_4$Sb$_4$ (ZrCoSb), Zr$_8$Co$_7$Ir$_1$Sb$_8$(ZrCo$_{0.875}$Ir$_{0.125}$Sb) and Zr$_4$Co$_3$Ir$_1$Sb$_4$ (ZrCo$_{0.75}$Ir$_{0.25}$Sb) are stable. For ZrCoSb, there are 12 atoms in the conventional unit cell which give rise to 36 phonon branches: three acoustic modes (1 longitudinal mode (LA) and 2 transverse modes (TA and TA')) and 33 optical phonon modes. Similarly, in ZrCo$_{0.75}$Ir$_{0.25}$Sb, there are 12 atoms leading to formation of 36 phonon branches (3 acoustic phonon modes (LA/TA/TA') and 33 optical phonon modes). However, in case of ZrCo$_{0.875}$Ir$_{0.125}$Sb, there are 24 atoms in the conventional unit cell, which leads to formation of 72 phonon branches (3 acoustic phonon modes and 69 optical phonon modes). Here, comparison has been done between the Zr$_4$Co$_4$Sb$_4$ and Zr$_4$Co$_3$Ir$_1$Sb$_4$ alloys. As the conventional unit cell of these alloys contains equal number of atoms, the number of optical and acoustic modes remains the same in both cases. However, Zr$_8$Co$_7$Ir$_1$Sb$_8$ consists of twice the number of atoms as compared with the former case, which can affect the magnitude of the $v_g$ and lowest frequency of optical phonon branches. Hence, comparison of the dispersion curve of $Zr_8Co_7Ir_1Sb_8$ with other two alloys can lead to wrong interpretation of the results. Fig. 3 (a) and (b) represents the phonon dispersion of $Zr_4Co_4Sb_4$, and $Zr_4Co_3Ir_1Sb_4$ alloys. It can be seen that the substitution of Co by Ir has induced significant changes in the phonon dispersion of the parent alloy. A significant overlap between acoustic branches and optical branches can also be observed. This is noted due to lowering of optical phonon modes due to substitution of heavier Ir atom at Co-site. The lowest frequency optical phonon branches located near Γ-X point move from ~ 2.89 THz ($Zr_4Co_4Sb_4$) to ~ 2.81 THz in $Zr_4Co_3Ir_1Sb_4$. It can lead to coupling between optical and acoustic phonon modes which can result in increment in Umklapp processes of phonon scattering due to the large number of phonon excitations in the vicinity of zone boundary. These characteristics are consistent with earlier reported phonon dispersion relations of other HHA such as ZrNiSn, ZrCoSb [31]. Additionally, a pronounced variation in the sound velocity of the acoustic phonon modes is observed. The calculated sound velocities in the vicinity of Γ for $Zr_4Co_4Sb_4$ and $Zr_4Co_3Ir_1Sb_4$ corresponding to TA/TA'/LA'/average acoustic phonon modes are 2.04/2.75/3.85/2.64 km/s and 1.98/2.5/3.74/2.47 km/s, respectively. There is decrement of 6.3% of $v_g$ in case of $Zr_4Co_3Ir_1Sb_4$ in comparison with $Zr_4Co_4Sb_4$, which indicates the softening of the acoustic phonon modes with increment in Ir concentration at Co-site. However, the experimentally noted reduction in $\kappa_L$ (~ 73%) cannot be alone accounted based on the softening of these modes and increment in Umklapp processes (due to coupling between acoustic and low frequency optical modes). It suggests that there is significant contribution from high frequency optical phonon modes in the observed magnitude of $\kappa_L$. Thus, in order to explain the drastic reduction in $\kappa_L$, point defect scattering which affects the high frequency optical phonon modes must be present in the system to explain this trend. The temperature dependent behaviour of $\kappa_L$ of these alloys is also analysed. From, eqn. (1) it can be inferred that $\kappa_L$ should exhibits $T^1$ behaviour due to dominance of three phonon scattering mechanism. As shown in Fig. 2 (c), the lattice thermal conductivity of ZrCoSb exhibits $T^1$ dependency throughout the measured temperature regime, however, in other two alloys, $T^1$ behaviour is noted above 750 K. This reflects the presence of different phonon mechanism below 750 K, which is discussed in the next paragraph.

The effect of point defect on the observed magnitude of RT $\kappa_L$ is analysed using the Klemen's–Callaway theory [32]. This model will help us to understand the similar magnitude of $\kappa_L$ in $ZrCo_{0.9}Ir_{0.1}Sb$, and $ZrCo_{0.75}Ir_{0.25}Sb$, by discerning the contribution from mass and strain field fluctuation effect. According to this theory, the ratio of lattice thermal conductivity of the alloyed ZrCo$_{1-x}$Ir$_x$Sb ($\kappa_0$) to that of ZrCoSb ($\kappa_0^P$) can be expressed in the form of $\frac{tan^{-1}u}{u} = \frac{\kappa_0}{\kappa_0^P}; u = (\frac{6\pi^2 V^2}{2k_B v_s})^{\frac{1}{3}}\kappa_0\Gamma$. Here it is assumed that ZrCoSb is the ordered crystal and Ir substituted alloys are disordered crystals and $u$ is the disorder parameter which depends on the physical properties of the ordered crystal; its lattice thermal conductivity ($\kappa_0$), average lattice sound velocity ($v_s$), average volume per atom as well as on scattering parameter ($\Gamma$). The obtained values of $\Gamma$ corresponding to each alloy is listed in the table 2. Here, $\Gamma$ is the summation of $\Gamma_{mass}$ and $\Gamma_{strain}$, where $\Gamma_{mass}$ is the mass fluctuation scattering parameter and $\Gamma_{strain}$ is the strain field fluctuation scattering parameter. The individual contributions from each component ($\Gamma_{mass}$ and $\Gamma_{strain}$) are theoretically estimated for ZrCo$_{0.9}$Ir$_{0.1}$Sb and ZrCo$_{0.75}$Ir$_{0.25}$Sb for Co sublattice using the following equations [33]

$\Gamma_{mass} = \frac{1}{3}[x(1-x)(\frac{M_{Co}-M_{Ir}}{M_{aver.}})^2]$ and $\Gamma_{strain} = \frac{1}{3}[\frac{(1-x)M_{Co}+xM_{Ir}}{M_{aver.}}]^2 x(1-x)\varepsilon[\frac{r_{Ir}-r_{Co}}{xr_{Ir}+(1-x)r_{Co}}]^2$

, where $M_{Co}$ and $M_{Ir}$ are the atomic mass of the Co and Ir, respectively, $M_{aver.}$ is the average atomic weight of the alloy, $r_{Ir}$ and $r_{Co}$ are the atomic radii of the Ir and Co respectively, x is the content of the Ir, $\varepsilon$ is the adjustable parameter for the Co sublattice, ranging from 10 to 100. The obtained parameters are listed in table 2. As noted from table 2, $\Gamma$ is comparable in ZrCo$_{0.9}$Ir$_{0.1}$Sb and ZrCo$_{0.75}$Ir$_{0.25}$Sb, which indicates that the influence of point defect scattering on both alloys is similar. Thus, as observed, no significant change in the magnitude of $\kappa_L$ of these alloys is noted. However, the contribution from $\Gamma_{mass}$ and $\Gamma_{strain}$ differs in both cases. In ZrCo$_{0.9}$Ir$_{0.1}$Sb, $\Gamma_{mass}$ is comparable with the $\Gamma_{strain}$; indicating equal contribution from mass fluctuation and strain field fluctuation scattering effect in the reduction of $\kappa_L$. However, in ZrCo$_{0.75}$Ir$_{0.25}$Sb, it is noted that $\Gamma_{mass}$ is significantly larger than $\Gamma_{strain}$. This signifies that as the concentration of Ir increases in the Co sublattice, mass fluctuation scattering plays a dominating role over strain field fluctuation scattering in reducing RT $\kappa_L$. From these observations, it can be said that in case of ZrCo$_{0.9}$Ir$_{0.1}$Sb the reduction in $\kappa_L$ can arise due to the mass fluctuations and strain field effects. However, in ZrCo$_{0.75}$Ir$_{0.25}$Sb it is noted due to presence of mass fluctuation effect. Hence, it can be concluded that partial substitution of Co by heavier element like Ir is an effective way to reduce the $\kappa_L$ in ZrCoSb-based compounds. Similar observations were noted in the TiNiSn, ZrNiSn and TiCoSb systems [34-36]. Additionally, in Ir substituted alloys, $\kappa_L$ shows $T^{-0.5}$ dependency below 500 K, which is indication of strong point defect scattering as described by Klemen's (as shown in the inset of Fig. 2 (c)) [17]. It can be said that in the measured temperature regime, Ir doped alloys display two different temperature dependent behaviour above and below bipolar regime. Below 500 K, there is dominance of point defect scattering mechanism whereas above 750 K, the thermal conductivity is governed by three phonon scattering process. This indicates that there are different mechanisms i.e., point defect scattering and phonon-phonon scattering responsible for heat transport in the Ir doped alloys. However, the reduction in RT $\kappa_L$ can be ascribed to the dominance of point defect scattering mechanism. In this particular case, the introduction of Ir atom at host atom Co-site results in changes in mass and strain field at the substituted site, as discussed above.

In summary, the structural and thermal transport properties of the $ZrCo_{1-x}Ir_xSb$ (x = 0, 0.1 and 0.25) HHA have been discussed in detail. Our analysis indicates that partial substitution of Ir at Co-site leads to softening of acoustic phonon modes and point defect scattering, which results in drastic suppression of $\kappa_L$ in $ZrCo_{0.9}Ir_{0.1}Sb$. No significant changes have been noted in $ZrCo_{0.75}Ir_{0.25}Sb$, because of identical scattering parameter. Interestingly, below 500 K, a dominance of point defect scattering mechanism is observed in Ir substituted alloys, whereas in ZrCoSb, phonon-phonon scattering plays a vital role in heat transport. Above 750 K, phonon-phonon scattering assisted thermal conductivity has been noted throughout this series. Our investigations might provide a pathway to researchers to effectively tune the large thermal conductivity of Zr based HHA through iso-electronic heavy mass substitution, for prospective thermoelectric applications.

KM acknowledges financial support from DST-SERB project CRG/2020/000073.

## DATA AVAILABILITY

The data that support the findings of this study are available from the corresponding author upon reasonable request.

**Tables**

**Table 1:** Obtained structural parameters from Rietveld refinement of $ZrCo_{1-x}Ir_xSb$ (x=0, 0.1 and 0.25)

| Alloys | ZrCoSb | $ZrCo_{0.9}Ir_{0.1}Sb$ | $ZrCo_{0.75}Ir_{0.25}Sb$ |
|---|---|---|---|
| Lattice parameter | 6.068(3) | 6.096(1) | 6.131(1) |
| Volume ($A^3$) | 223.46(3) | 226.55(1) | 230.569(1) |

**Table 2:** Disorder scaling parameter (u), disorder scattering parameters ($\Gamma_{expt.}$, $\Gamma_{mass}$, $\Gamma_{stra}$) and the strain related adjustable parameter for the Co sublattice ($\varepsilon$) for $ZrCo_{1-x}Ir_xSb$ (x=0, 0.1 and 0.25)

|  | u | $\Gamma_{expt.}$ | $\Gamma_{mass}$ | $\Gamma_{stra}$ | $\varepsilon$ |
|---|---|---|---|---|---|
| **ZrCoSb** | - | - | - | - | - |
| **$ZrCo_{0.9}Ir_{0.1}Sb$** | 4.88 | 0.097 | 0.050 | 0.047 | 55 |
| **$ZrCo_{0.75}Ir_{0.25}Sb$** | 5.04 | 0.119 | 0.104 | 0.01 | 10 |

**Figures**

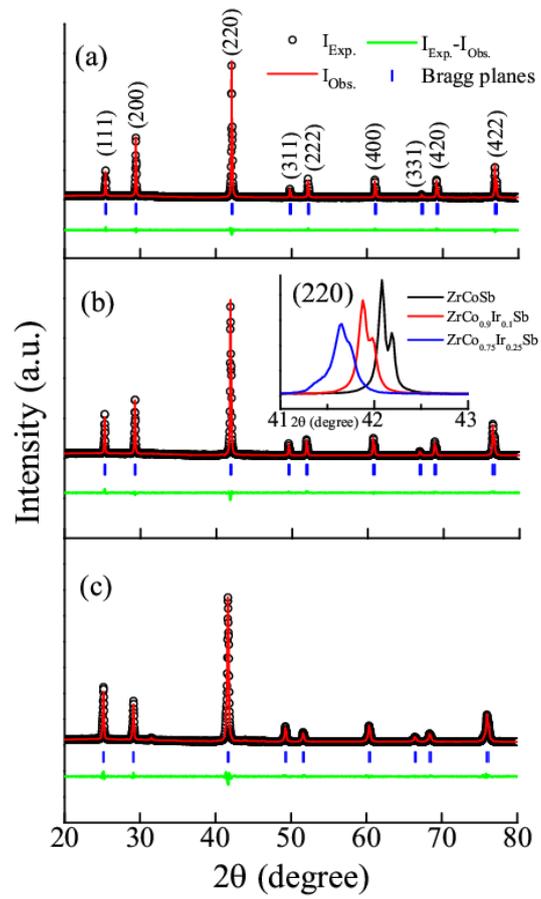

**Fig. 1** Rietveld refined RT XRD patterns of (a) ZrCoSb (b) ZrCo$_{0.9}$Ir$_{0.1}$Sb (c) ZrCo$_{0.75}$Ir$_{0.25}$Sb alloys Inset of (b): shows the shifting of (220) peak towards lower angle side with increment in Ir concentration.

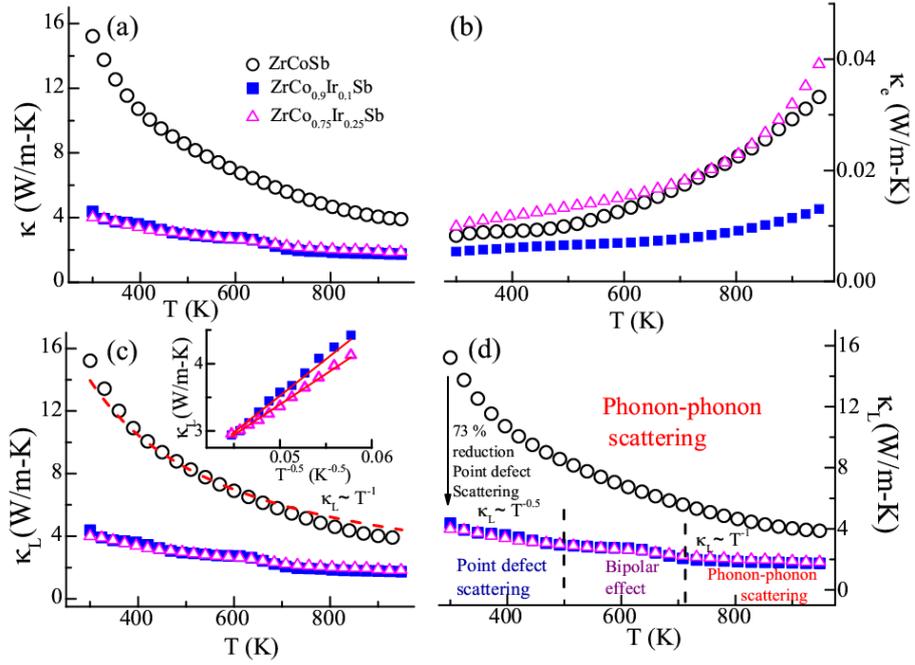

**Fig. 2** Temperature response of (a) $\kappa$ and (b) $\kappa_e$ in the temperature range 300-950 K (c) $\kappa_L$ vs T plot in the same range; Red dashed line represents the $T^{-1}$ dependency of $\kappa_L$ Inset: $\kappa_L$ vs $T^{-0.5}$ plot; Red solid line represents linear fitting corresponding to $ZrCo_{1-x}Ir_xSb$ (x = 0, 0.1 and 0.25) HHA (d) Illustration to demonstrate the dominance of different phonon scattering mechanism in the measured temperature range.

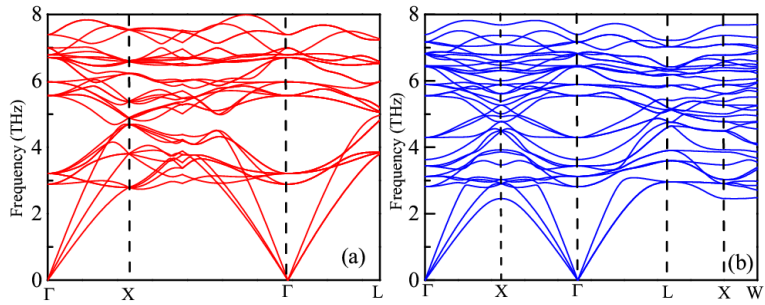

**Fig. 3** Phonon dispersion curve of (a) $Zr_4Co_4Sb_4$ and (b) $Zr_4Co_3Ir_1Sb_4$ along high symmetric *k*-points

# Supplementary material for

# Unravelling the phonon scattering mechanisms in Half-Heusler ZrCo$_{1-x}$Ir$_x$Sb (x= 0, 0.1, 0.25) alloys

Kavita Yadav[1*], Saurabh Singh[2], Omprakash Muthuswamy[2], Tsunehiro Takeuchi[2], and K. Mukherjee[1]

[1]School of Basic Sciences, Indian Institute of Technology, Mandi, Himachal Pradesh-175005, India

[2]Research Centre for Smart Energy Technology, Toyota Technological Institute, Nagoya, 468-8511, Japan

**Table S1** Obtained parameters from Rietveld refinement of ZrCo$_{1-x}$Ir$_x$Sb (0≤x≤1) alloys

| Alloy | ZrCoSb | | | | |
|---|---|---|---|---|---|
| Bragg R factor | 1.78 | | | | |
| Rf factor | 1.29 | | | | |
| $\chi^2$ | 1.97 | | | | |
| Wyckoff positions | X | Y | Z | position | Occupancy |
| Zr | 0 | 0 | 0 | 4a | 1.0 |
| Co | 0.25 | 0.25 | 0.25 | 4c | 1.0 |
| Sb | 0.50 | 0.50 | 0.50 | 4b | 1.0 |
| Vacant sites | 0.75 | 0.75 | 0.75 | 4d | 0.0 |
| Alloy | ZrCo$_{0.9}$Ir$_{0.1}$Sb | | | | |
| Bragg R factor | 2.47 | | | | |
| Rf factor | 1.31 | | | | |
| $\chi^2$ | 1.71 | | | | |
| Wyckoff positions | X | Y | Z | position | Occupancy |
| Zr | 0 | 0 | 0 | 4a | 1.0 |
| Co | 0.25 | 0.25 | 0.25 | 4c | 0.9 |
| Ir | 0.25 | 0.25 | 0.25 | 4c | 0.1 |
| Sb | 0.50 | 0.50 | 0.50 | 4b | 1.0 |
| Vacant sites | 0.75 | 0.75 | 0.75 | 4d | 0.0 |
| Alloy | ZrCo$_{0.75}$Ir$_{0.25}$Sb | | | | |
| Bragg R factor | 1.56 | | | | |
| Rf factor | 0.86 | | | | |
| $\chi^2$ | 2.70 | | | | |
| Wyckoff positions | X | Y | Z | position | Occupancy |
| Zr | 0 | 0 | 0 | 4a | 1.0 |
| Co | 0.25 | 0.25 | 0.25 | 4c | 0.75 |
| **Ir** | 0.25 | 0.25 | 0.25 | **4c** | 0.25 |
| **Sb** | 0.50 | 0.50 | 0.50 | **4b** | 1.0 |
| **Vacant sites** | 0.75 | 0.75 | 0.75 | **4d** | 0.0 |

**Table S2:** Comparison between theoretical and experimental intensities of fundamental peaks of $ZrCoSb$, $ZrCo_{0.9}Ir_{0.1}Sb$ and $ZrCo_{0.75}Ir_{0.25}Sb$

| **ZrCoSb** | **Theoretical intensity** | **Experimental Intensity** |
|---|---|---|
| **111** | 9.28 | 13.72 |
| **200** | 50.25 | 56.47 |
| **220** | 100 | 100 |
| $I_{111}/I_{200}$ | 0.18 | 0.24 |
| $I_{200}/I_{220}$ | 0.50 | 0.56 |
| **$ZrCo_{0.9}Ir_{0.1}Sb$** | | |
| **111** | 20.87 | 11.64 |
| **200** | 31.41 | 35.98 |
| **220** | 100 | 100 |
| $I_{111}/I_{200}$ | 0.66 | 0.32 |
| $I_{200}/I_{220}$ | 0.31 | 0.35 |
| **$ZrCo_{0.75}Ir_{0.25}Sb$** | | |
| **111** | 15.36 | 27.48 |
| **200** | 29.10 | 20.86 |
| **220** | 100 | 100 |
| $I_{111}/I_{200}$ | 0.52 | 1.31 |
| $I_{200}/I_{220}$ | 0.29 | 0.20 |

**Figure S1** Temperature response of 'S' of $ZrCo_{1-x}Ir_xSb$ (x= 0, 0.1 and 0.25) alloys measured in the 300-950 K temperature range.

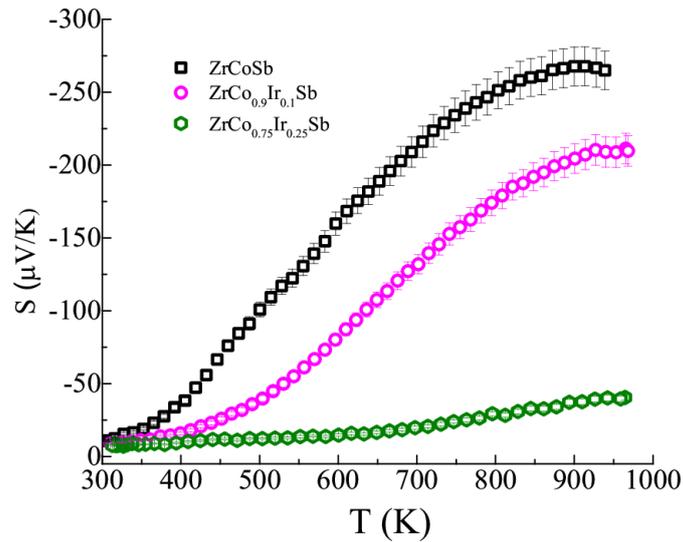